\documentstyle[amssymb,prl,aps,epsfig]{revtex}

\begin{document}

\title{Evidence for a smooth superconductor to normal state transition
for nonzero applied magnetic field in RbOs$_{2}$O$_{6}$}

\author{T. Schneider$^{\text{1}}$, R. Khasanov$^{\text{1,2,3}}$, J. Karpinski$^{\text{4}}$, S. M. Kazakov$^{\text{4}}$ and H. Keller$^{\text{1}}$}
\address{$^{\text{(1)}}$ Physik-Institut der Universit\"{a}t Z\"{u}rich,\\Winterthurerstrasse 190, CH-8057, Switzerland\\
$^{\text{(2)}}$ Laboratory for Neutron Scattering, ETH Z\"{u}rich and PSI\\Villigen, CH-5232 Villigen PSI, Switzerland\\
$^{\text{(3)}}$DPMC, Universit\'e de Gen\`eve, 24 Quai Ernest-Ansermet, 1211 Gen\`eve 4, Switzerland\\$^{\text{(4)}}$ Laboratory for Solid State\\
Physics, ETH Z\"{u}rich, 8093, Switzerland}
\maketitle

\begin{abstract}
The effect of the magnetic field on the critical behavior of
RbOs$_{2}$O$_{6} $ is investigated near the zero field transition
at $T_{c}$. We present and analyze magnetization data, revealing
for $0.01\leq \mu _{0}H\leq 1$ T remarkable consistency with a
magnetic field induced finite size effect. It is traced back to
the fact that at temperatures $T_{p}\left( H\right) <T_{c}$ the
correlation length cannot grow beyond the limiting length set by
the magnetic field. Thus, for nonzero $H$ the transition from the
superconducting to the normal state turns out to be smooth and the
appropriately scaled magnetization data fall on a universal curve.
\end{abstract}
\bigskip

We investigate the effect of the magnetic field on the critical
behavior of the recently discovered superconductor
RbOs$_{2}$O$_{6}$\cite{yonezawarb} near the zero field transition
at $T_{c}$ by taking thermal fluctuations into account. This
compound is particularly suited because near $T_{c}$ charged
critical fluctuations have been shown to dominate\cite{tsrkhk},
while the low temperature properties are well described by the BCS
approximation\cite{khasanov,khasanov2}. Furthermore,
RbOs$_{2}$O$_{6}$ appears to be a nearly isotropic superconductor.
Traditionally magnetization data have been interpreted in terms of
\cite{abrikosov}
\begin{equation}
-4\pi M\left( T\right) =\frac{H_{c2}-H}{\left( 2\kappa
^{2}-1\right) \beta _{A}},  \label{eq1}
\end{equation}
where $\beta _{A}=1.16$ for a hexagonal vortex lattice. $\kappa
=\lambda /\xi $ denotes the Ginzburg-Landau parameter, $\xi =\xi
_{0}\left( 1-T/T_{c0}\right) ^{-1/2}$ the correlation length ,
$\lambda =\lambda _{0}\left( 1-T/T_{c0}\right) ^{-1/2}$ the
magnetic penetration depth, $T_{c0}$ the mean-field transition
temperature, and $H_{c2}=\Phi _{0}/\left( 2\pi \xi ^{2}\right)
$the so called upper critical field. Thus, the magnetization
vanishes as $H\rightarrow H_{c2}$, because of the assumed
continuous phase transition, where $\xi $ diverges. At fixed field
the temperature dependence of the magnetization is the given by
\begin{equation}
4\pi M\left( T\right) =-\frac{1}{\left( 2\kappa ^{2}-1\right)
\beta _{A}}\left( \frac{\Phi _{0}}{2\pi \xi _{0}^{2}}\left(
1-\frac{T}{T_{c0}}\right) -H\right) ,  \label{eq2}
\end{equation}
and
\begin{equation}
4\pi \frac{dM\left( T\right) }{dT}=\frac{1}{\left( 2\kappa
^{2}-1\right) \beta _{A}}\frac{\Phi _{0}}{2\pi \xi
_{0}^{2}T_{c0}}.  \label{eq3}
\end{equation}
Because this mean-field treatment neglects thermal fluctuations it
fails whenever these fluctuations dominate.

In this study we present and analyze extended field cooled
ac-magnetization measurements. Invoking the scaling theory of
critical phenomena it is shown that the data are inconsistent with
the aforementioned mean-field prediction for $0.01\leq \mu
_{0}H\leq 1$ T. On the contrary, we observe agreement with a
magnetic field induced finite size effect. Indeed, when the
magnetic field increases, the density of vortex lines becomes
greater, but this cannot continue indefinitely, the limit is
roughly set on the proximity of vortex lines by the overlapping of
their cores. Because of the resulting limiting length scale
$L_{H}$ the correlation length $\xi $ cannot grow beyond\cite
{haussmann,lortz,parks,bled}
\begin{equation}
L_{H}=\sqrt{\Phi _{0}/aH},  \label{eq4}
\end{equation}
with $a\simeq 3.12$\cite{bled}. It is comparable to the average
distance between vortex lines and implies that $\xi $ cannot grow
beyond $L_{H}$ and with that there is a magnetic field induced
finite size effect. This implies that thermodynamic quantities
like the magnetization, magnetic penetration depth, specific heat
\textit{etc.} are smooth functions of temperature near
$T_{p}\left( H\right) $, where the correlation length cannot grow
beyond $\xi \left( T\right) =L_{H}$. This scenario holds true when
the magnetization data $M\left( T,H\right) $ collapses near
$T_{p}\left( H\right) $ on a single curve when plotted as
$M/(TH^{1/2})$ vs. $\left( T/T_{c}-1\right) /\left( 1-T_{p}\left(
H\right) /T_{c}\right)$, where $T_{c}$ is the zero field
transition temperature. For $0.01\leq \mu _{0}H\leq 1$ T we
observe that our magnetization data falls within experimental
error on a single curve by adjusting $T_{p}\left( H\right) $. From
the resulting field dependence of $T_{p}$ we deduce for the
critical amplitude of the correlation length the estimate $\xi
_{0}\simeq 74$\AA . Thus, although RbOs$_{2}$O$_{6}$ exhibits BCS
ground state properties\cite{khasanov,khasanov2}, thermal
fluctuations do not alter the zero field thermodynamic properties
near $T_{c}$ only\cite{tsrkhk}, but invalidate the assumption of
an upper critical field $H_{c2}$ over the rather extended field
range $0.01\leq H\leq 1$ T. As a result, whenever the thermal
fluctuation dominated regime is accessible, type II
superconductors in a nonzero magnetic field do not undergo a
continuous phase transition, e.g. to a state with zero resistance.

Polycrystalline samples of RbOs$_{2}$O$_{6}$ were synthesized by a
procedure similar to that described by Yonezawa \textit{et
al.}\cite{Yonezawa04}, Kazakov \textit{et al.}\cite{Kazakov04} and
Br\"{u}hwiler \textit{et al.}\cite {bruhwiler}. The field-cooled
ac-magnetization measurements were performed with a commercial
PPMS (Physical Property Measurement System) magnetometer in a
fields of 0~T-8~T and at temperatures ranging from 1.75~K to 10~K.
The ac-frequency was set to 82~Hz and the amplitude to 0.5~mT.

\begin{figure}[tbp]
\centering
\includegraphics[totalheight=6cm]{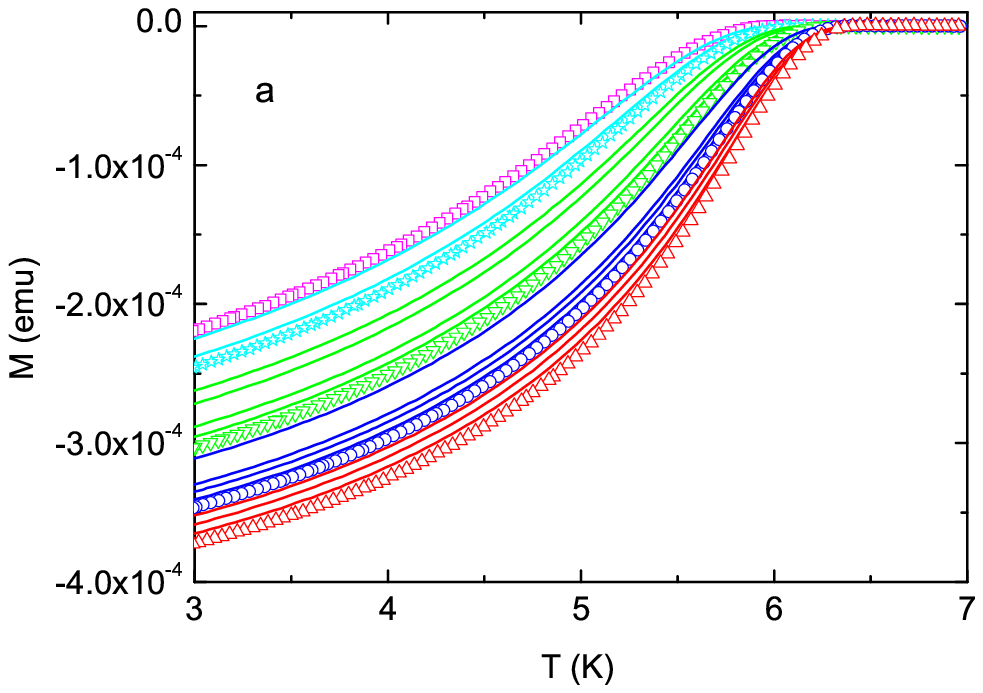}
\includegraphics[totalheight=6cm]{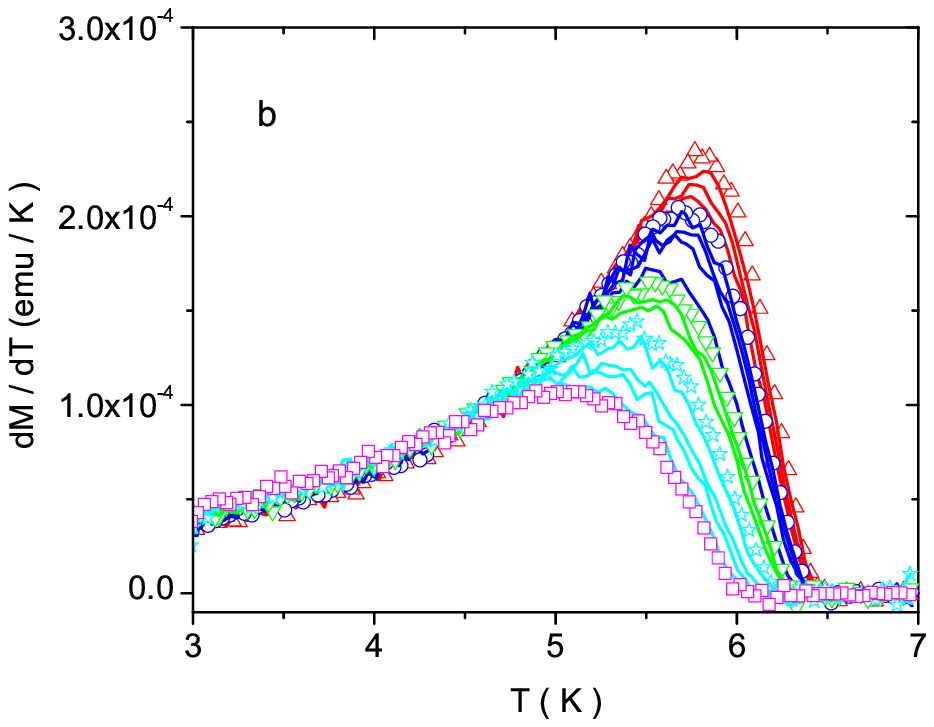}
\caption{a) Field cooled magnetization $M$ of a RbOs$_{2}$O$_{6}$
powder sample \textit{vs}. $T$ for various applied magnetic fields
($\mu _{0}H=0.11(\triangle )$, $0.12$, $0.13$, $0.14$,
$0.15(\bigcirc )$, $0.16$, $0.17 $, $0.18$, $0.22$,
$0.24(\triangledown )$, $0.26$, $0.28$, $0.333(\star )$, $0.366$,
$0.433$, $0.466$, and 0.533 T ($\square $); b) $dM/dT$
\textit{vs.} $T$ for the data shown in Fig.\ref{fig1}a.}
\label{fig1}
\end{figure}

In Fig.\ref{fig1}a we displayed some low field data in terms of
$M$ \textit{vs.} $T$ and in Fig.\ref{fig1}b the respective $dM/dT$
\textit{vs.} $T$. To identify the temperature regime where
critical fluctuations play an essential role it is instructive to
consider $dM/dT$ \textit{vs.} $T$. With decreasing temperature
$dM/dT$ is seen to raise below the zero field transition
temperature $T_{c}\simeq 6.5$ K and after passing a maximum value
it decreases. Thus, except for the temperature region around the
maximum, $dM/dT$ does not adopt a constant value as the mean-field
treatment, neglecting thermal fluctuations, suggests
(Eq.(\ref{eq3})). Indeed, the gradual raise of $dM/dT$ below
$T_{c}\simeq 6.5$ K uncovers the effect of thermal fluctuations
and indicates that in contradiction to the mean-field
approximation there is no sharp transition in an applied magnetic
field.

\begin{figure}[tbp]
\centering
\includegraphics[totalheight=6cm]{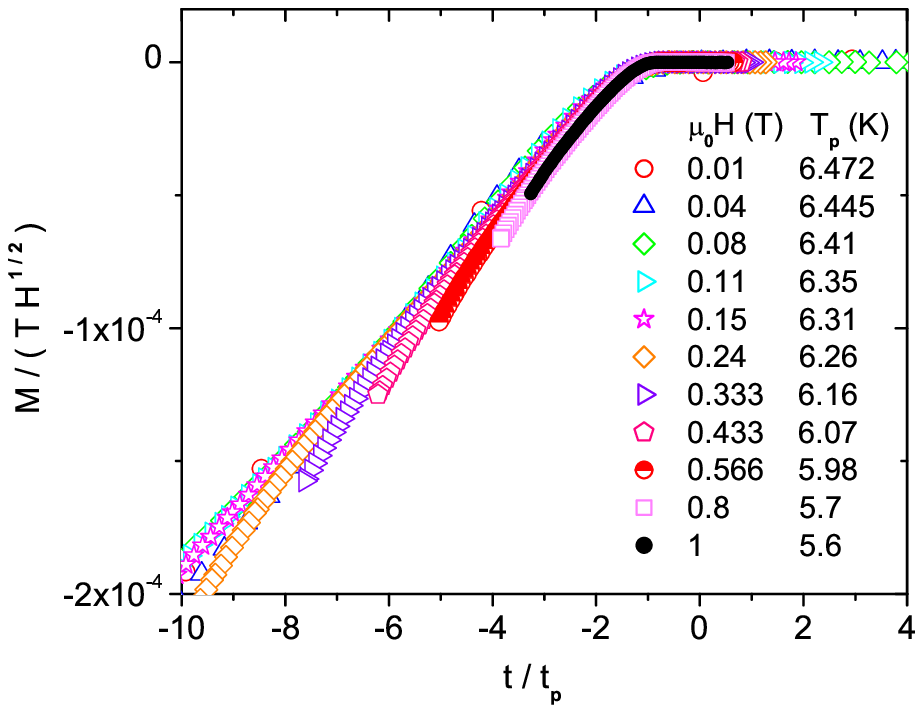}
\caption{$M/(TH^{1/2})$ vs. $t/t_{p}\left( H\right) =\left(
T/T_{c}-1\right) /\left( 1-T_{p}\left( H\right) /T_{c}\right) $
for various magnetic fields with $T_{c}=6.5$K and the listed
estimates for $T_{p}\left( H\right) $} \label{fig2}
\end{figure}

When the rounding of the transition stems from a magnetic field or
inhomogeneity induced finite size effect, the correlation length
$\xi $ cannot grow beyond the limiting length $L_{H,I}$, where
\begin{equation}
\xi \left( T_{p}\right) =\xi _{0}\left| t_{p}\right| ^{-\nu
}=L_{H,I},\text{ }t=1-T_{p}/T_{c},\nu \simeq 2/3.  \label{eq5}
\end{equation}
$L_{I}$ denotes the limiting length of the homogeneous domains of
the sample. Note that $\nu \simeq 2/3$ holds in the charged
universality class as well\cite{tsrkhk}.In superconductors,
exposed to a magnetic field $H$ , there is the aforementioned
additional limiting length scale $L_{H}=\sqrt{\Phi _{0}/\left(
aH\right) }$(Eq.(\ref{eq4})), related to the average distance
between vortex lines. Indeed, as the magnetic field increases, the
density of vortex lines becomes greater, but this cannot continue
indefinitely, the limit is roughly set on the proximity of vortex
lines by the overlapping of their cores. Because of these limiting
length scales the phase transition is rounded and occurs smoothly.
Consequently, the thermodynamic quantities like the magnetization,
magnetic penetration depth, specific heat \textit{etc.} are smooth
functions of temperature near $T_{p}$. To uncover the scaling
properties of the magnetization in this regime and to estimate the
magnetic field dependence of $T_{p}$, we invoke the scaling
properties of the free energy per unit volume in the regime where
the thermal fluctuations dominate. Although the order parameter
fluctuations are coupled to fluctuations in the vector potential,
in an applied magnetic field the order parameter fluctuations
dominate\cite{brezin}. In this case the free energy per unit
volume adopts the scaling form\cite {parks,bled,lawrie,book}
\begin{equation}
f=\frac{Qk_{B}T}{\xi ^{3}}G\left( z\right),~z=\frac{H\xi
^{2}}{\Phi _{0}}, \label{eq7}
\end{equation}
where $Q$ is a universal constant and $G\left( z\right) $ a
universal function of its argument. For the magnetization
$m=-df/dH$ we obtain then the scaling relation
\begin{equation}
\frac{m}{T\sqrt{H}}=-\frac{Qk_{B}}{\Phi _{0}^{3/2}}F\left(
z\right) ,~F\left( z\right) =z^{-1/2}\frac{dG\left( z\right)
}{dz}. \label{eq8}
\end{equation}
Thus, when thermal fluctuations dominate and the magnetic field
induced finite size effect sets the limiting length
($L_{H}<L_{I}$), data as shown in Fig.\ref{fig1}a should collapse
on a single curve when plotted as $M/(TH^{1/2})$ \textit{vs.}
$t/t_{p}\left( H\right) =\left( T/T_{c}-1\right) /\left(
1-T_{p}\left( H\right) /T_{c}\right) $ with appropriately chosen
zero field $T_{c}$ and $T_{p}\left( H\right) $. Indeed, according
to Eqs.(\ref{eq5}),  (\ref{eq7}) and the critical behavior of the
zero field correlation length $\xi \left( T\right) =\xi _{0}\left|
t\right| ^{-\nu }$ with $t=T/T_{c}-1$ the scaling variable $z$ can
be expressed as $z^{-1/2\nu }=a^{1/2\nu }t/t_{p}\left( H\right) $.
A glance to Fig.\ref{fig2} shows that for $T_{c}=6.5$K and the
listed values of $T_{p}\left( H\right) $ the data tends to
collapse around $t/t_{p}\left( H\right) =-1$ on a single curve.
However, with increasing field the scaling regime where the data
collapse is seen to shrink. This reflects the fact that the
fluctuations of a bulk superconductor in sufficiently high
magnetic fields become effectively one dimensional, as noted by
Lee and Shenoy\cite{lee}. Here a bulk superconductor behaves like
an array of rods parallel to the magnetic field with diameter
$L_{H}$, while the scaling relation (\ref{eq8}) holds for
sufficiently low fields where three dimensional fluctuations
dominate. On the other hand, with decreasing magnetic field the
scaling regime is seen to increase. Thus, down to $0.01$T the
magnetic field sets the limiting length scale so that
$L_{I}>L_{H=0.01T}\simeq 2575$\AA . Accordingly, we established
the consistency with a magnetic field induced finite size effect,
revealing that in a magnetic field RbOs$_{2}$O$_{6}$ does not
undergo a sharp phase transition from the superconducting to the
normal state up to at least $1$T. In principle, the scaling
function $G\left( z\right) $ should also have a singularity at
some value $z_{m}$ of the scaling variable below $T_{p}\left(
H\right) $ describing the vortex melting transition, but this
singularity is not addressed here.

\begin{figure}[tbp]
\centering
\includegraphics[totalheight=6cm]{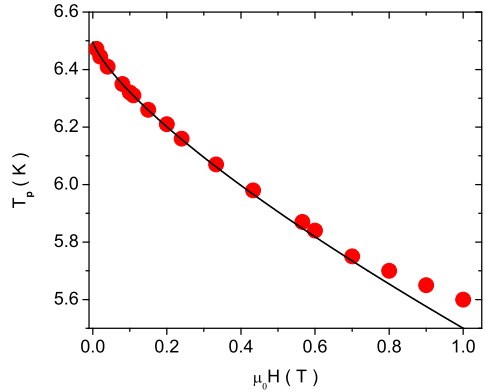}
\caption{$T_{p}$ \textit{vs.} $H$ derived from the scaling plots
shown in Fig.\ref{fig2}. The solid line is Eq.(\ref{eq9}) with
$T_{c}=6.5$K , $\left( a\xi _{0}^{2}/\Phi _{0}\right)
^{3/4}=0.154$, $a=3.12$ and $\nu =2/3$.} \label{fig3}
\end{figure}

In Fig.\ref{fig3} we displayed our estimates for $T_{p}\left(
H\right) $, the temperature where the correlation length equals
the magnetic field induced limiting length scale. According to
Eqs.(\ref{eq4}) and (\ref{eq5}) the leading field dependence is
\begin{equation}
T_{p}\left( H\right) =T_{c}\left( 1-\left( \frac{aH\xi
_{0}^{2}}{\Phi _{0}}\right) ^{1/2\nu }\right) .  \label{eq9}
\end{equation}
The solid line in Fig.\ref{fig3} is this relation with
$T_{c}=6.5$K , $\left( a\xi _{0}^{2}/\Phi _{0}\right)
^{3/4}=0.154$, $a=3.12$ and $\nu =2/3$, yielding for the critical
amplitude of the correlation length the estimate $\xi _{0}\simeq
74$\AA. This value agrees reasonably well with $\xi _{0}\simeq
84$\AA \cite{tsrkhk} derived from the magnetic field induced shift
of the specific heat peak\cite{bruhwiler}. Since Eq.(\ref{eq9})
describes the leading behavior in the limit $H\rightarrow 0$ the
occurrence of systematic deviations with increasing magnetic field
point to the aforementioned magnetic field induced dimensional
crossover\cite{lee}.

We have shown that even in RbOs$_{2}$O$_{6}$, exhibiting BCS
ground state properties\cite{khasanov,khasanov2}, thermal
fluctuations do not alter the zero field thermodynamic properties
near $T_{c}$ only\cite{tsrkhk}, but invalidate the assumption of a
continuous phase transition at an upper critical field
$H_{c2}\left( T\right) $ over a rather extended temperature range.
We observed a rounded transition. It was traced back to a magnetic
field induced finite size whereupon the correlation length cannot
grow beyond the limiting length $L_{H}=\sqrt{\Phi _{0}/aH}$,
comparable to the average distance between vortex lines. As a
result and in agreement with numerical
studies\cite{kajantiea,nguyen}, whenever the thermal fluctuation
dominated regime is accessible, there is in type II
superconductors no critical line $H_{c2}\left( T\right) $ of
continuous phase transitions.

\acknowledgments This work was partially supported by the Swiss
National Science Foundation and the NCCR program {\it Materials
with Novel Electronic Properties} (MaNEP) sponsored by the Swiss
National Science Foundation.

\end{document}